\documentclass[aps,amsmath,amssymb,floatfix,twocolumn,prl]{revtex4-1}
\usepackage[]{standalone}
\usepackage{times}
\usepackage{graphicx}
\usepackage{color}
\usepackage{bm}
\usepackage{braket}
\usepackage{mathtools}
\usepackage{mathcomp}
\usepackage[caption=false,justification=raggedright,position=top,singlelinecheck=off]{subfig}
\usepackage{soul}
\usepackage[mathscr]{euscript}

\newcommand{\enni}{\noindent}
\newcommand{\enbe}{\begin{equation}}
\newcommand{\enee}{\end{equation}}

\begin{document}
\title{Landau levels from neutral 
Bogoliubov particles in two-dimensional nodal superconductors under 
strain and doping gradients}

\author{Emilian M.\ Nica}
\email[Corresponding author: ]{enica@qmi.ubc.ca}

\author{Marcel Franz}
\affiliation{Department of Physics and Astronomy and Quantum Materials Institute,
University of British Columbia, Vancouver, B.C., V6T 1Z1, Canada}

\date{\today}

\begin{abstract}
Motivated by recent work on strain-induced pseudo-magnetic fields 
in Dirac and Weyl semimetals, 
we analyze the possibility of analogous fields in
two-dimensional nodal superconductors. We consider the prototypical case of a 
$d$-wave superconductor, a representative of the cuprate family, and 
find that the presence of weak strain leads to pseudo-magnetic fields and 
Landau quantization of Bogoliubov quasiparticles in the low-energy
sector. A similar effect is  induced by the presence of generic, weak doping gradients. 
In contrast to genuine magnetic fields in superconductors, 
the strain- and doping gradient-induced pseudo-magnetic 
fields couple in a way that preserves time-reversal symmetry 
and is not subject to the screening associated with the Meissner effect. 
These effects can be probed 
by tuning weak applied supercurrents which lead
to shifts in the energies of the Landau levels and 
hence to quantum oscillations in thermodynamic 
and transport quantities.  
\end{abstract}

\maketitle
\emph{Introduction.--}
Elementary excitations in superconductors are comprised of coherent
superpositions of electron and hole degrees of freedom
\cite{BCS,dG,Tinkham_1996}. These Bogoliubov quasiparticles are electrically neutral
on average and therefore do not couple simply to the externally
applied magnetic field. In addition, superconductors are known to
expel magnetic field from their bulk either completely
\cite{Meissner}, or form a flux lattice \cite{Abrikosov}, in which the
quasiparticle dynamics is effectively zero-field \cite{FT}. For
these reasons superconductors normally avoid formation of Landau levels
which represent the canonical response of most other electron systems to magnetic
field \cite{Schoenberg}.  

In this work, we show that weak, in-plane strains and doping gradients
generically 
lead to Landau quantization of Bogoliubov quasiparticles for a 
broad class of 2D nodal superconductors (SC). 
In these cases, the Dirac-like quasiparticles in the vicinity 
of point nodes are subject to emergent vector potentials
which enter in a time-reversal invariant way. In contrast 
to genuine magnetic fields in a SC, there are no 
induced currents and no screening associated with the Meissner effect.
Our work is motivated by interesting developments in graphene
\cite{Guinea_Nat_Phys_2009} where strain-induced pseudo-magnetic fields
lead to Landau quantization and quantum oscillations, 
which were already observed in experiment~\cite{Levy_Science_2010}, 
and more recent proposals in  the context of Dirac and Weyl semimetals
~\cite{Shapourian,Cortijo,Fujimoto,Pikulin,Grushin,Liu_PRB_2017}. 

A possible experimental setup is shown in Fig.~\ref{fig1}. 
Since the most immediate realization of gapless, effectively
two-dimensional superconductivity is provided by 
the broad class of Cu-based materials, 
we study the case of a prototypical $d$-wave SC. Strain 
can be induced in principle by allowing for the controlled 
deformation of an underlying substrate \cite{Bozovic,AChen2016,Guzman2017}. Controlled doping gradients~\cite{Taylor2015, Wu2013}
provide an alternate way of introducing unconventional vector potentials.
As discussed in greater detail in the following, an additional application 
of a weak supercurrent provides a simple way to detect
the underlying quantum oscillations.  
\noindent \begin{figure}[t!]
\includegraphics[width=7cm]{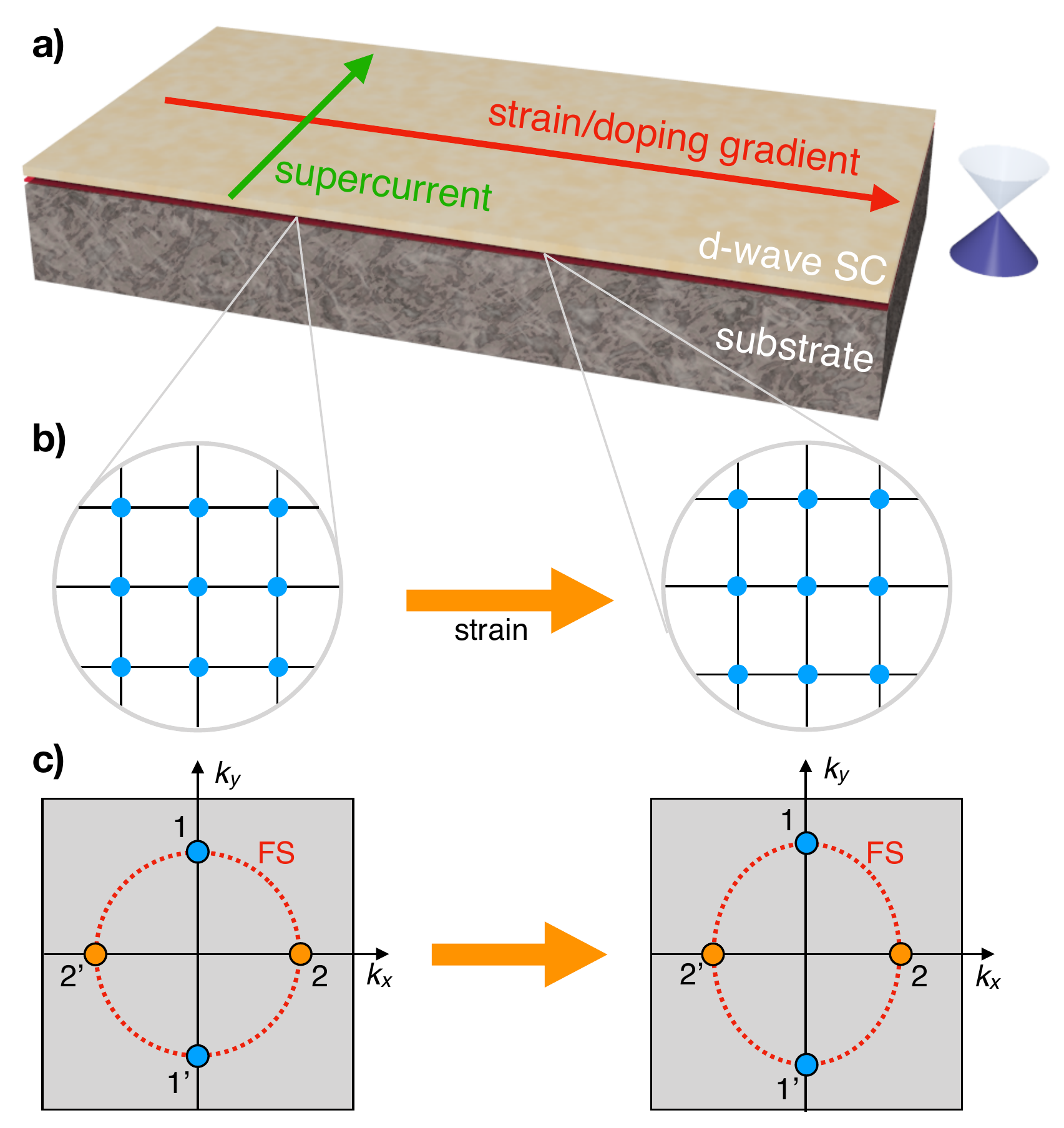}
\caption{a) Sketch of a possible 
experimental setup. The red arrow indicates
the direction of variation of either strain 
or doping gradient. The green arrow
indicates an applied supercurrent. 
Panel (b)
shows the schematic deformation of 
the 2D square lattice for the case of 
uniaxial strain  as a function of $x$ alone.
The corresponding Fermi surface (c) deforms and the nodal points,
labeled (1,1') and (2,2'), move, mimicking the action of a
pseudo-vector potential.
}
\label{fig1}
\end{figure}

\emph{Vector potentials from lattice deformations and doping
  gradients. --}

Consider
a prototypical Hamiltonian \cite{Simon1997} on the square lattice
$\hat{H}=\hat{H}_{TB} + \hat{H}_{\Delta}$
\enni corresponding to a nearest-neighbor (NN) tight-binding (TB)
part, together with a pairing potential at mean-field level for 
next-nearest neighbors (NNN). The latter is chosen to belong
to the $d_{xy}$ representation of the $D_{4h}$ point group. 
In momentum space, the Nambu form of the lattice Hamiltonian is
written as $\hat{H} = \sum_{\bm{k}}  \mathbf{\Psi}^{\dagger}_{\bm k}
\mathscr{H}(\bm{k}) \mathbf{\Psi}_{\bm k}$ with $\Psi_{\bm k}=(c_{\bm{k}, \uparrow}, c_{\bm{k}, \downarrow}, c^{\dagger}_{\bm{-k}, \uparrow}, c^{\dagger}_{\bm{-k}, \downarrow})^{T}$ 
the Nambu spinor and 
\enni \enbe
\mathscr{H}(\bm{k}) = h_{\bm k} \sigma_{0} \tau_{z} +\Delta_{\bm k} (i\sigma_{y}) (i \tau_{y}).
\enee 
Here ${\bm \sigma}$ and ${\bm \tau}$ are Pauli matrices
in spin and Nambu space, respectively, $h_{\bm k} = 2t[ \cos(k_{x}a) +
\cos(k_{y}a)] - \mu$ and 
$\Delta_{\bm k} = 4 \Delta \sin(k_{x}a) \sin(k_{y}a)$.  In addition,
$t, \Delta$ are the unperturbed hopping coefficients and pairing amplitudes, 
respectively, and $a$ is the pristine NN lattice spacing. 

In the absence of any perturbation and 
below half-filling, the low-energy spectrum of $\hat{H}$ is Dirac-like
\enni \enbe
E^{(\alpha)}_{\bm q}= \pm \sqrt{v^{2}_{F} q^{2}_{x/y} + v^{2}_{\Delta} q^{2}_{y/x}},
\enee 
 about four nodes located at $\bm{K}_{\alpha} \in \{(\pm K_{F}, 0), (0, \pm K_{F}) \}$, 
where $K_{F}$ is the Fermi wavevector. We label the pairs of opposite 
momenta as $\alpha \in\{1, 1'\}$ for the nodes along the $k_{y}$ axis and 
 $\alpha \in\{2, 2'\}$ for the nodes along the $k_{x}$ axis in the Brillouin Zone, as
shown in Fig.~\ref{fig1} (c).
We also define the effective Fermi velocities 
$v_{F}= 2ta \sin(K_{F}a), v_{\Delta} = 4 \Delta a \sin(K_{F}a)$.

We model an arbitrary lattice deformation 
via the transformation 
\enni \enbe
\label{Eq:Trns_lttc}
\bm{R}_{i} \rightarrow \bm{R}^{'}_{i} = \bm{R}_{i} + \bm{\epsilon}(\bm{R}_{i}),
\enee
\enni where $\bm{R}_{i}$ are Bravais lattice vectors and 
$\bm{\epsilon}(\bm{R}_{i})$ are position-dependent displacements of the orbitals. 
We assume that both NN hopping coefficients and NNN pairing amplitudes
are continuous functions of the deformation. While in practice 
they can be quite sensitive to the details of the material at hand, 
we assume that the effect of a deformation can be generically modeled
by considering the leading contributions in a gradient expansion 
of a deformation field $\bm{\epsilon}(\bm{r})$. 
In addition, we also assume that the leading effect 
can be captured by a net change 
in NN bond length, by analogy with the case
of graphene~\cite{Suzuura_PRB_2002, Vozmediano_Phys_Rep_2010}. 
We also ignore the  
contributions from the change in pairing, which are 
expected to be sub-leading. 
While we expect that none of these approximations are  
crucial for the study of the effect at hand, they provide 
for a much more transparent discussion. 

The pairing potential connects states in the vicinity of 
pairs of opposite Fermi wavevectors. Consequently, we can focus
on the $(1, 1')$ pair of nodes. 
\begin{figure*}[ht!]
 \subfloat{\includegraphics[width=0.6\columnwidth]{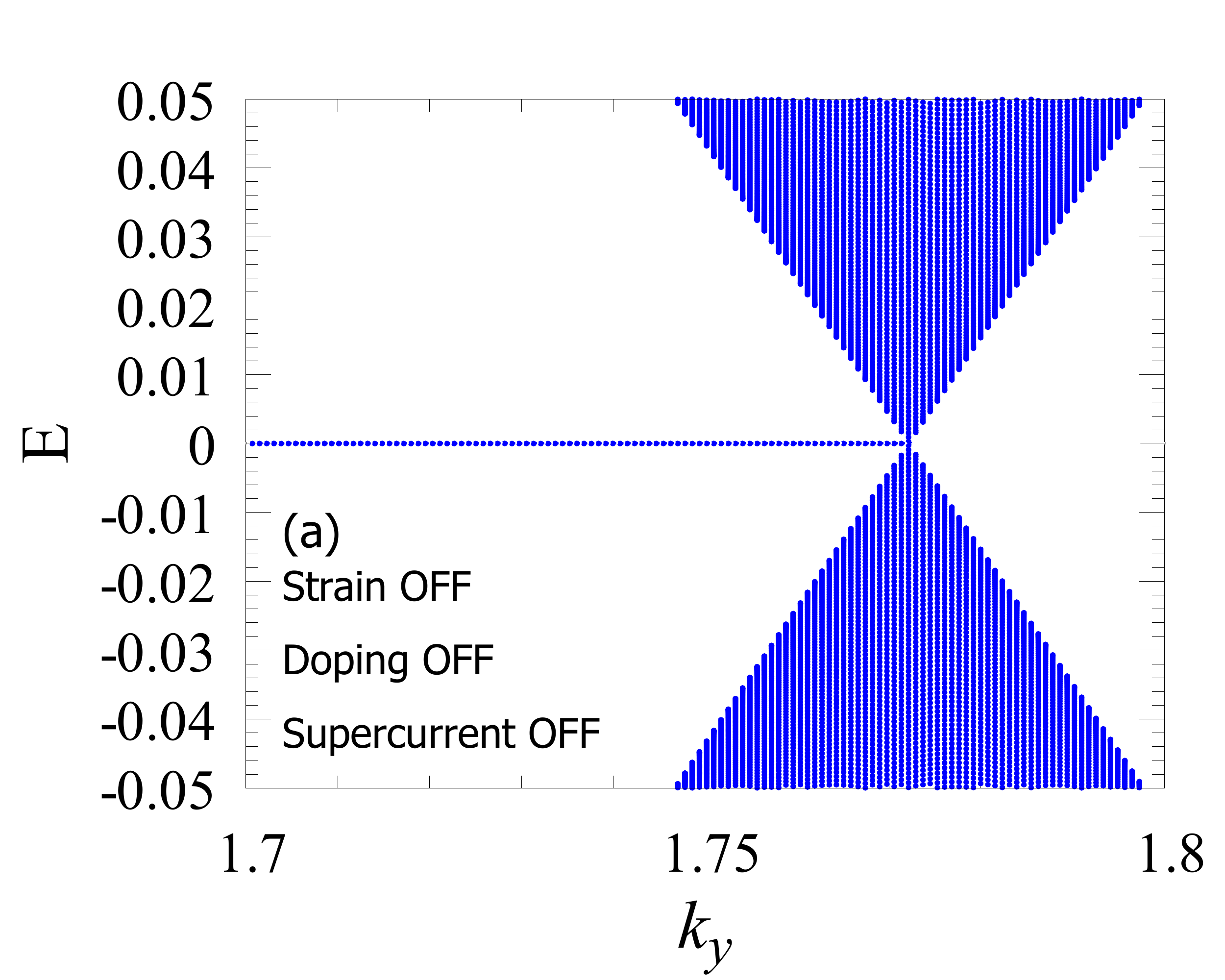}} \quad
\subfloat{\includegraphics[width=0.6\columnwidth]{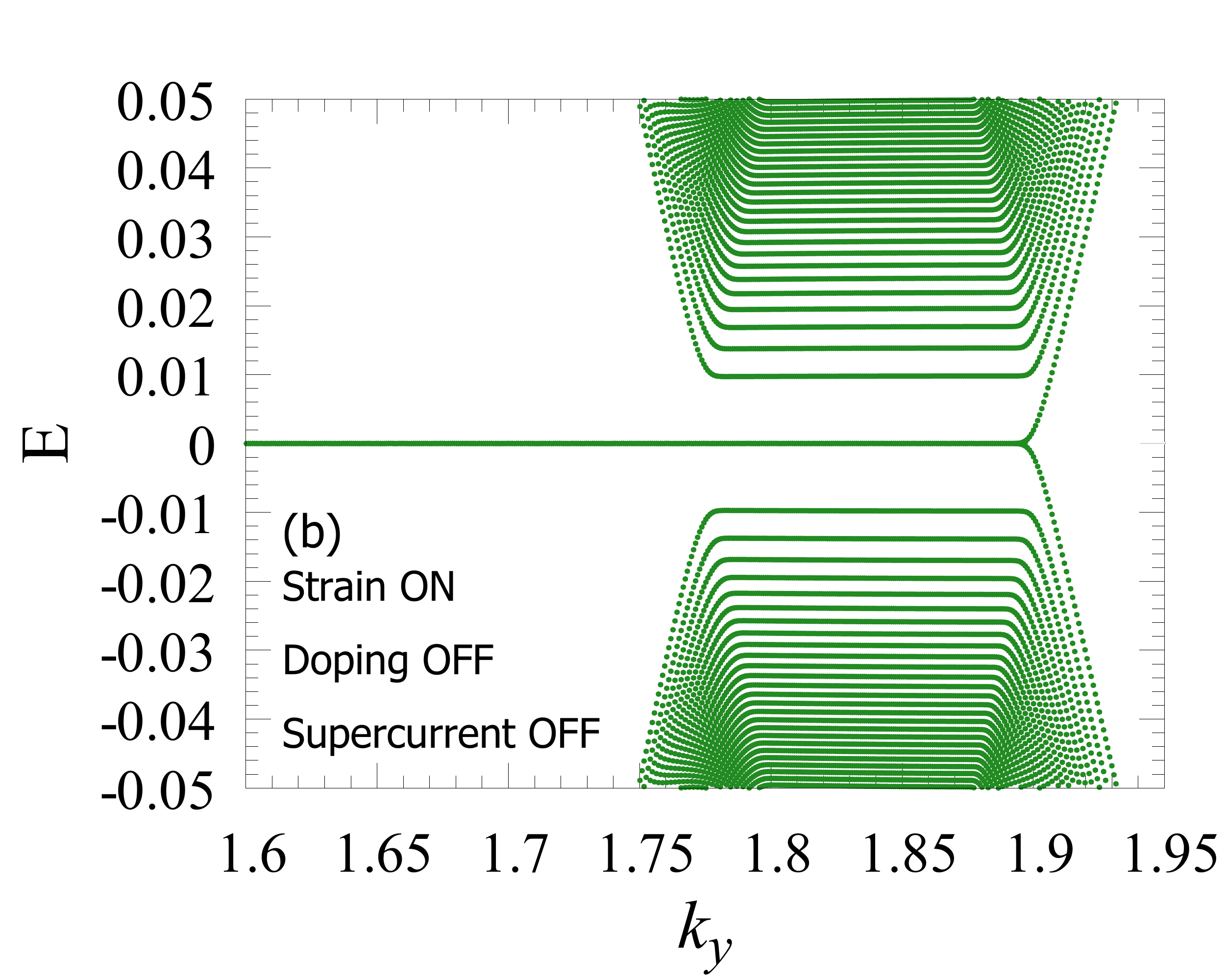}} \quad
 \subfloat{\includegraphics[width=0.6\columnwidth]{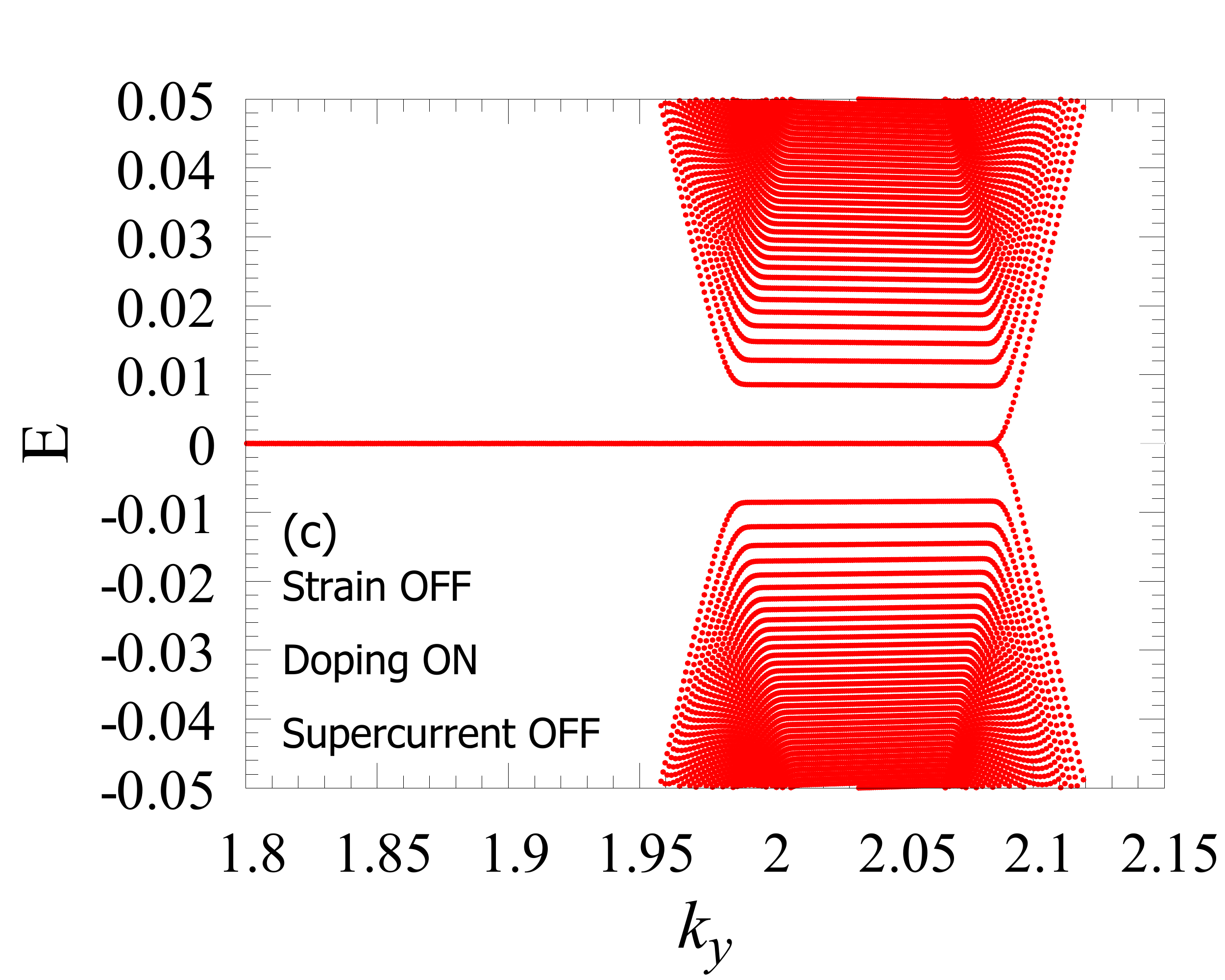}} \\[-3ex]
\subfloat{\includegraphics[width=0.6\columnwidth]{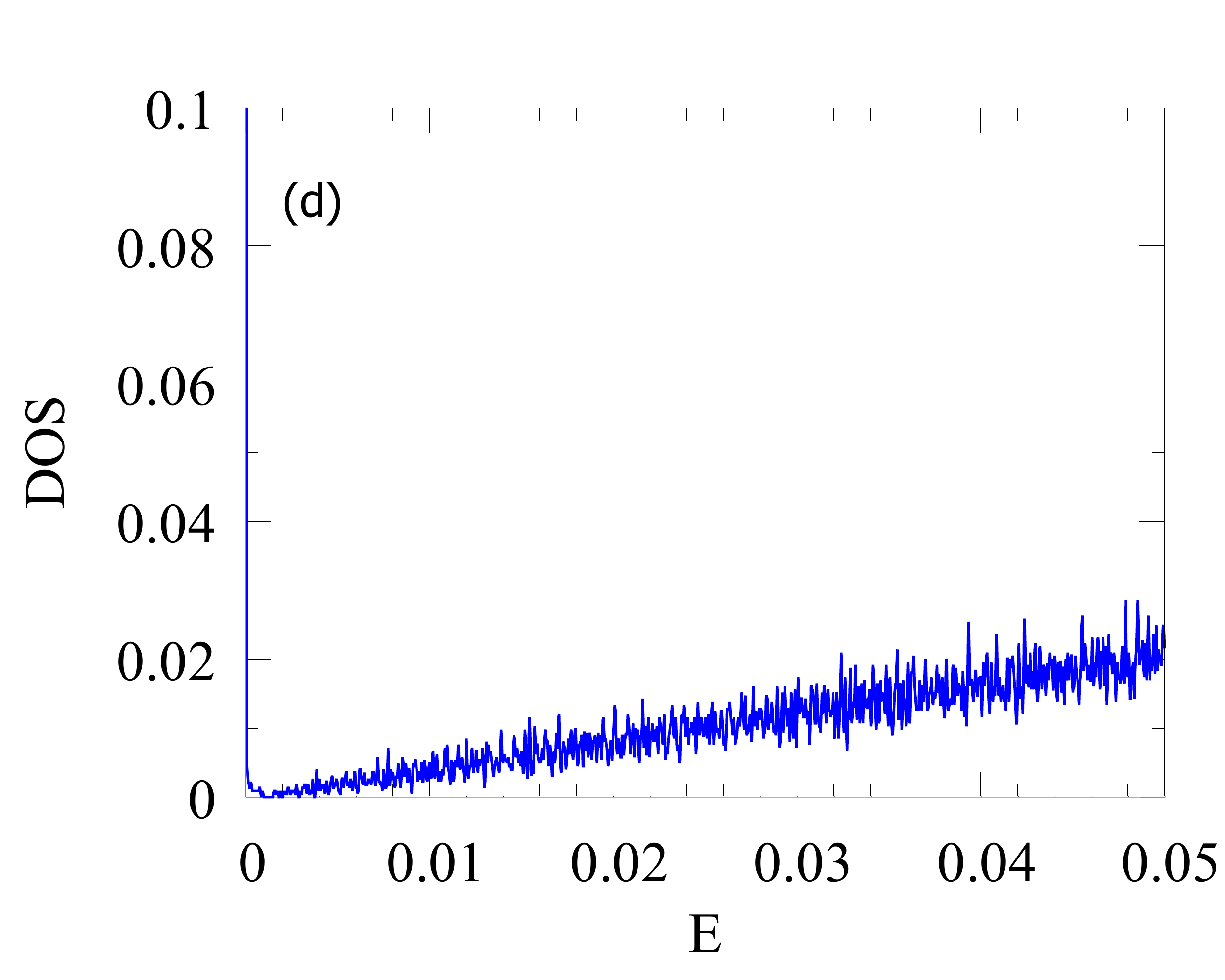}} \quad
 \subfloat{\includegraphics[width=0.6\columnwidth]{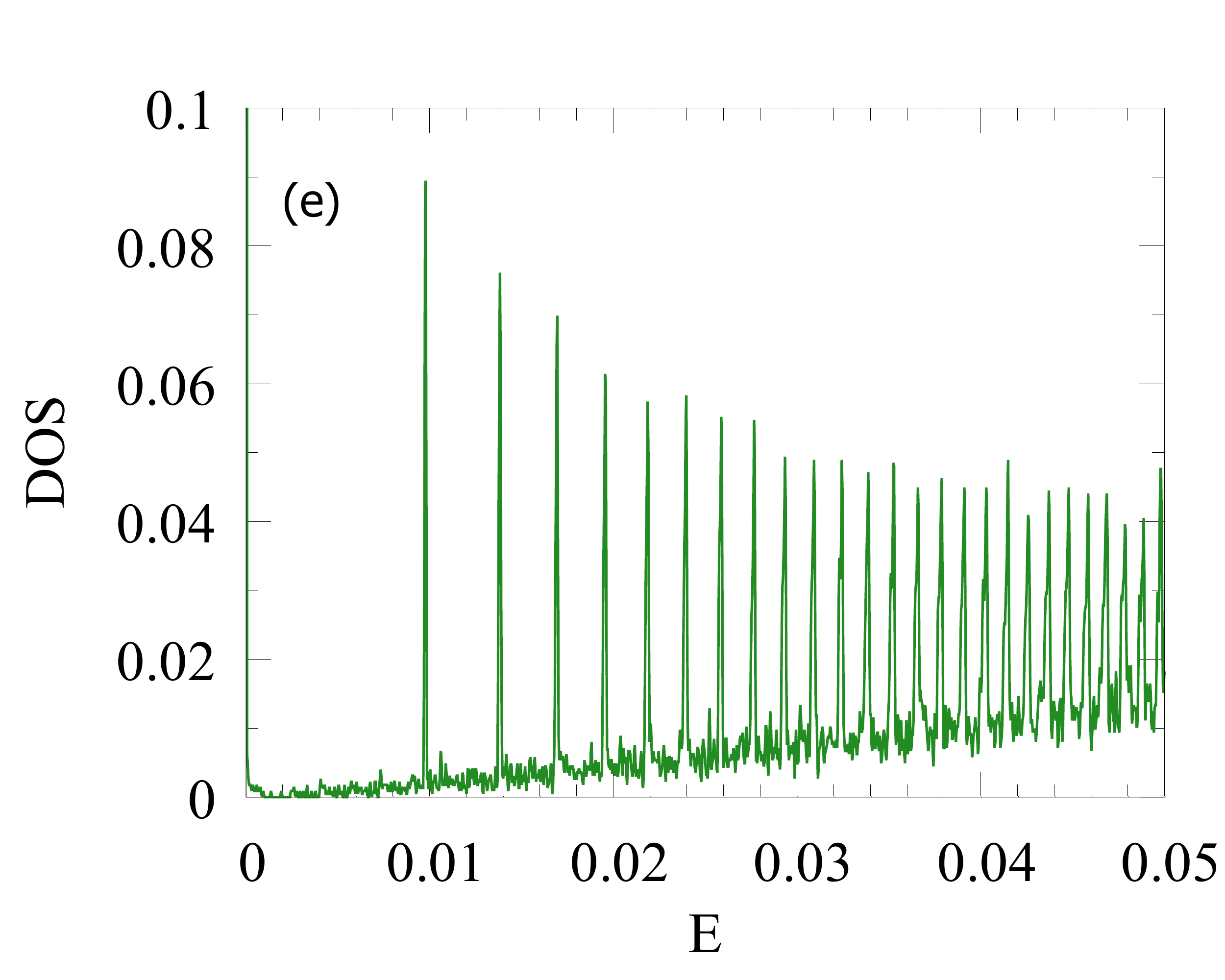}} \quad
\subfloat{\includegraphics[width=0.6\columnwidth]{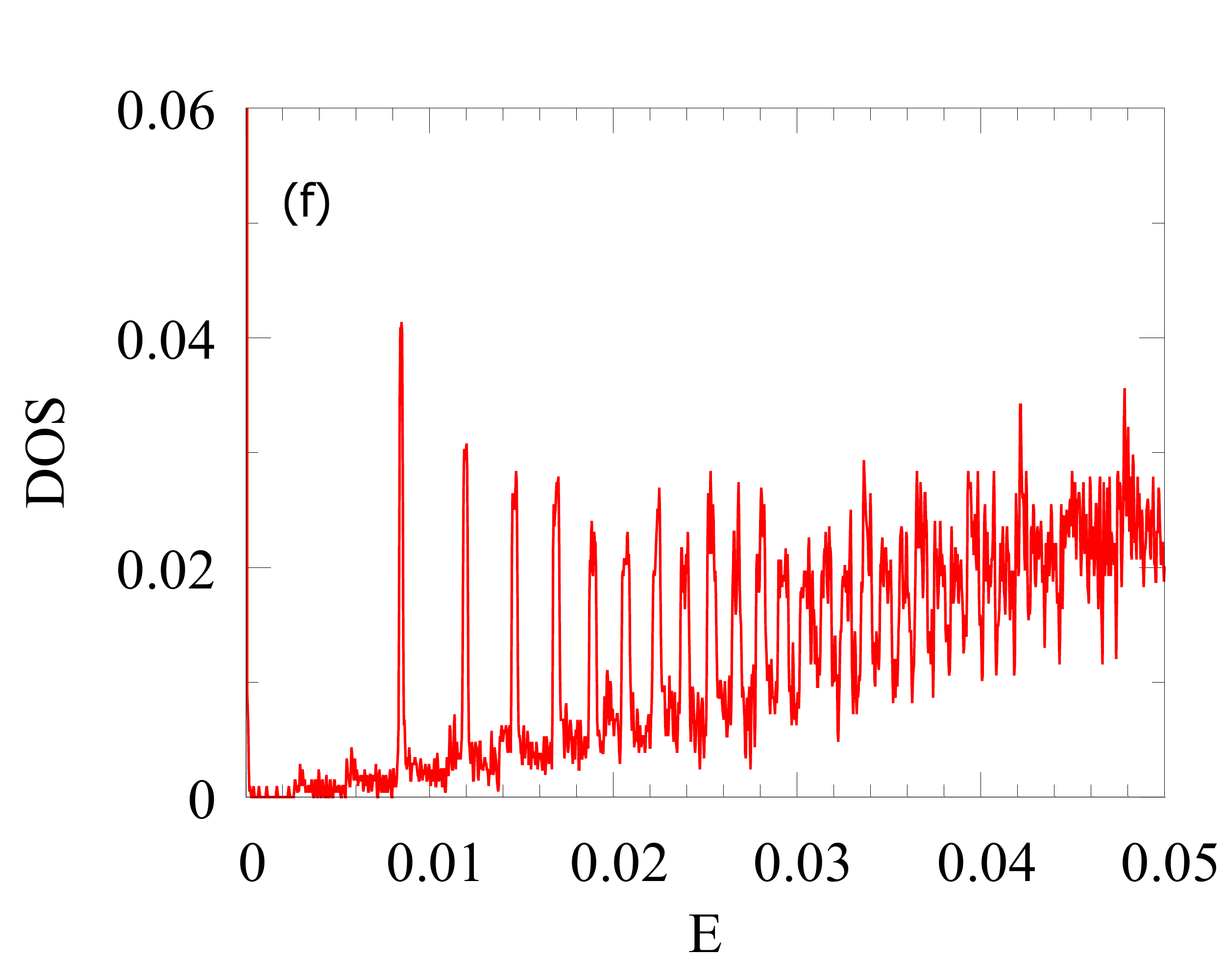}}
\caption{Results of the numerical calculations for 
the 2D $d_{xy}$ SC with and without the presence of strain and 
doping gradients.
Computations were performed on a lattice with $L_x=2000$
sites and $L_y=8000$ crystal momentum points. We use $\Delta=0.1$,
and $\mu=-1.6$ and $\mu=-1.0$ for
the strain and doping cases, respectively. 
(a) Low-energy spectrum around node 1
in the absence of either strain or doping gradients.
(b) Same as in (a) with a finite pure-shear strain with
$\delta_{sp}=5 \times 10^{-5}$  which 
varies along the $x$-direction.  (c) Same as in (b) 
with a doping gradient $\delta_{dp}=1 \times 10^{-4}$ instead of strain. (d)-(f) 
The total DOS per unit area for the cases in (a)-(c) respectively.
}
\label{Fig:Zr_crnt}
\end{figure*}
In the low-energy, continuum limit, the Hamiltonian 
reduces to $H^{(1, 1')} =  \int d^{2}r \mathbf{\Psi}^{\dagger}_{\bm
  r} \mathscr{H} \mathbf{\Psi}_{\bm r}$,
with 
\enni \begin{align}
\label{Eq:Hmlt_Nmb}
\mathscr{H} = v_{F}  \left( \sigma_{z} \tau_{0}  i\partial_{y} + \sigma_{0} \tau_{z} \frac{e\mathscr{A}_{y}}{v_{F}} \right) 
- v_{\Delta} \sigma_{x} \tau_{x} i\partial_{x} , 
\end{align}
\enni where $\hbar$ has been set to 1 for simplicity. 
A detailed derivation of this Hamiltonian is provided 
in the Supplementary Material (SM) but the origin of the vector
potential can be understood intuitively by inspecting Fig.\ \ref{fig1}(b,c).
Note that ${\bm \sigma}$
are now Pauli matrices 
in \emph{combined} valley and spin 
space, and
$\mathbf{\Psi}_{\bm{r}}= (\Psi^{(1)}_{\uparrow}, \Psi^{(1')}_{\downarrow}, \Psi^{(1), \dagger}_{\uparrow}, \Psi^{(1'), \dagger}_{\downarrow})$
is the corresponding Nambu spinor. The Fermi fields are defined for the 
pristine system in the vicinity of the nodes in standard fashion.
Notice that both kinetic and pairing parts are effectively 
one-dimensional in this limit. Consequently, 
the deformation-induced potentials, which couple 
in a gauge-invariant way, are of the form $\bm{\mathscr{A}}=(0, \mathscr{A}_{y})$. 
The effective one-dimensional form
also precludes the emergence of  
scalar potentials for a generic deformation, 
in contrast to the case of graphene
~\cite{Guinea_Nat_Phys_2009, Manes_PRB_2007, Manes_PRB_2013}, where this holds 
only for pure shear deformations. 
The effective Hamiltonian about the 
other two Fermi wavevectors can be obtained by transforming 
$x \leftrightarrow y$. 

Under our assumptions, the generic form of the vector 
potentials are (SM)
\enni \begin{align}
\label{Eq:Strn_gg}
\mathscr{A}_{y} = \left( \frac{2t \beta}{e} \right) \big[ u_{xx} + \cos(K_{F}a) u_{yy} \big],
\end{align}
\enni where $u_{ij}= (1/2)(\partial_{j}\epsilon_{i}+\partial_{i}\epsilon_{j})$ is a 
symmetric strain tensor and $\beta=d \ln t/ d \ln a$ is a standard parameter~\cite{Vozmediano_Phys_Rep_2010}.
Quite generally, the elements of the strain tensor
can be continuous functions of $(x,y)$. We list  
three limiting cases, which are 
more conveniently achieved, in numerical 
calculations and possible experimental setups:
(i) \emph{Uniaxial strain}, $u_{xx} \neq 0, u_{yy}=0$; 
(ii) \emph{Hydrostatic compression/dilation}, $u_{xx}=u_{yy}$; 
and (iii) \emph{Pure shear strain}, $u_{xx}=-u_{yy}$. 
In the following, we shall focus on case (iii), although 
this does not essentially modify the results.

A very similar form is obtained for 
the case of a doping gradient in the low-energy limit. 
This possibility can be modeled by introducing a slow variation 
of the chemical potential on the scale of the inter-site 
separation. In the low-energy, continuum
limit we approximate $\mu \rightarrow \mu(1+ g(\bm{r}))$.
As for the case of lattice deformations, the additional term
leads to the emergence of vector potentials of the form
\enni \begin{align}
\label{Eq:Dpng_grdn}
\mathscr{A}_{y} = \left( \frac{\mu}{e} \right) g(\bm{r}).
\end{align}
This again can be understood intuitively by noting that doping
gradient changes the Fermi surface volume and thus moves the nodal
points in the momentum space.

The vector potentials in Eqs.~\ref{Eq:Strn_gg} and~\ref{Eq:Dpng_grdn}
can lead to pseudo-magnetic fields, provided that
$\bm{\mathscr{B}} = \bm{\nabla} \times \bm{\mathscr{A}} \neq 0$. 
In practice, we analyze the case of strains with monotonic
linear variation with distance along $x$
and analogously for the chemical potential case. 
This corresponds to uniform
pseudo-magnetic fields. To ensure that the continuum limit 
is a good approximation to the perturbed finite-size system, 
we require that the components of 
the vector potentials remain small over 
the entire sample.

The vector potential terms  are invariant under the time-reversal 
operation, which effectively interchanges the valley and spin 
indices of the paired fields.
Consequently, the Hamiltonian in Eq.~\ref{Eq:Hmlt_Nmb} is also invariant 
under time-reversal. This ensures that the current density 
associated with either the deformation or doping gradient 
vanishes. 
An important consequence of this is the absence of screening currents and hence
of the Meissner effect, which would otherwise prevent 
the emergence of Landau levels (LL). 
We also note that, from a generic Landau-Ginzburg 
perspective, there is no analog
for the standard London equations as the 
effective strain-induced vector potentials are not 
determined from the standard gauge-invariant action. 
Instead, they are derived from 
a linear-elastic theory, as noted in the 
case of graphene~\cite{Vozmediano_Phys_Rep_2010}. 
Similar arguments hold for doping gradient-induced vector 
potentials. 

The solutions to Eq.~\ref{Eq:Hmlt_Nmb} can be 
obtained via a canonical Bogoliubov-de Gennes (BdG) 
transformation~\cite{dG}. 
For node 1 the BdG equations are
\enni \begin{align}
 - \bigg( v_{\Delta} i\partial_{x} \sigma_{x} +  v_{F} \left( i\partial_{y} + \frac{e\mathscr{A}_{y}}{v_{F}} \right) \sigma_{y} \bigg) \bm{\psi} = E \bm{\psi},
\label{Eq:BdG}
\end{align}
\enni where $\bm{\sigma}$ are Pauli matrices and 
$\bm{\psi}= (u^{(1)}_E(\bm{r}), v^{(1)}_{E}(\bm{r}))^{T}$ 
is a spinor associated with the BdG factors. For $\mathscr{A}_{y}$
linearly increasing along $x$ 
the eigenstates are discrete Landau levels of 
energy $E_{n}= \pm \omega_{c} \sqrt{n}$, 
where $\omega_{c}= \sqrt{2 e v_{\Delta} \partial_{x} \mathscr{A}_{y}(x)}$. 
The eigenstates at 1' are  obtained via complex-conjugation. 

For \emph{bona fide} magnetic fields it is well-known~\cite{Schoenberg} 
that Landau quantization generically 
leads to oscillations in thermodynamic and 
transport observables with applied field, most notably 
the de Haas-van Alphen
and Shubnikov-de Haas effects. 
Similar effects have been predicted for 
strain-induced pseudo-magnetic fields in Dirac~\cite{Guinea_Nat_Phys_2009} 
and Weyl materials~\cite{Liu_PRB_2017}. These oscillations
can be traced~\cite{Schoenberg} to the dramatic 
enhancements in the total density of states (DOS)
per unit volume whenever a LL crosses the Fermi energy. 
We argue that similar singularities arise in the present case,
and that the associated oscillations can 
be observed in principle in a suitable experimental setup. 
In the case
of a clean superconductor under strain or doping gradient, a convenient 
way to probe the discrete nature of the LL's is to apply 
a small supercurrent to the sample as indicated in Fig.\ \ref{fig1} (a). In the low-energy limit, 
the supercurrents induce an effective Doppler shift in quasiparticle energy~\cite{Tinkham_1996}
given by  $E_{\bm k}\to E_{\bm k} +{\bm v}_s\cdot {\bm k}$ where 
$\bm{v}_{s}=\bm{q}_{s}/m$ defines the superfluid velocity~\cite{dG}. 
This perturbation breaks time-reversal symmetry and, importantly, 
leads to opposite shifts around opposite momenta. 
We argue that this effect also occurs in the presence of 
strains or doping gradients.
For fixed effective vector potentials, the DOS at the Fermi level
will therefore exhibit sharp enhancements as a function of a weak, applied supercurrent.
Similar effects are expected in the presence of fixed supercurrents 
and varying strain or doping gradients. 
In the following, we present numerical calculations
which fully support our predictions. 
\begin{figure*}[ht!]
\subfloat{\includegraphics[width=0.6\columnwidth]{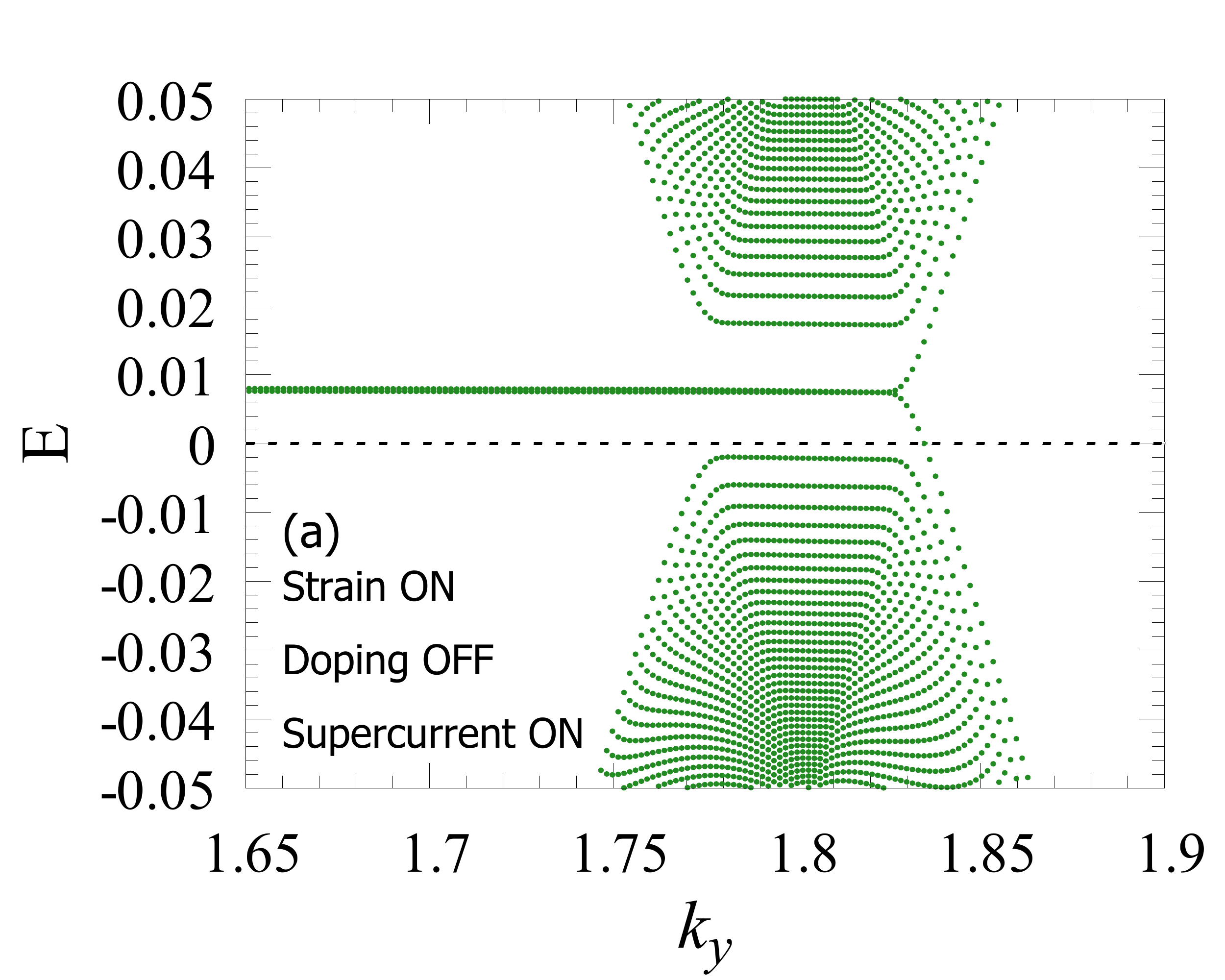}} \quad
 \subfloat{\includegraphics[width=0.6\columnwidth]{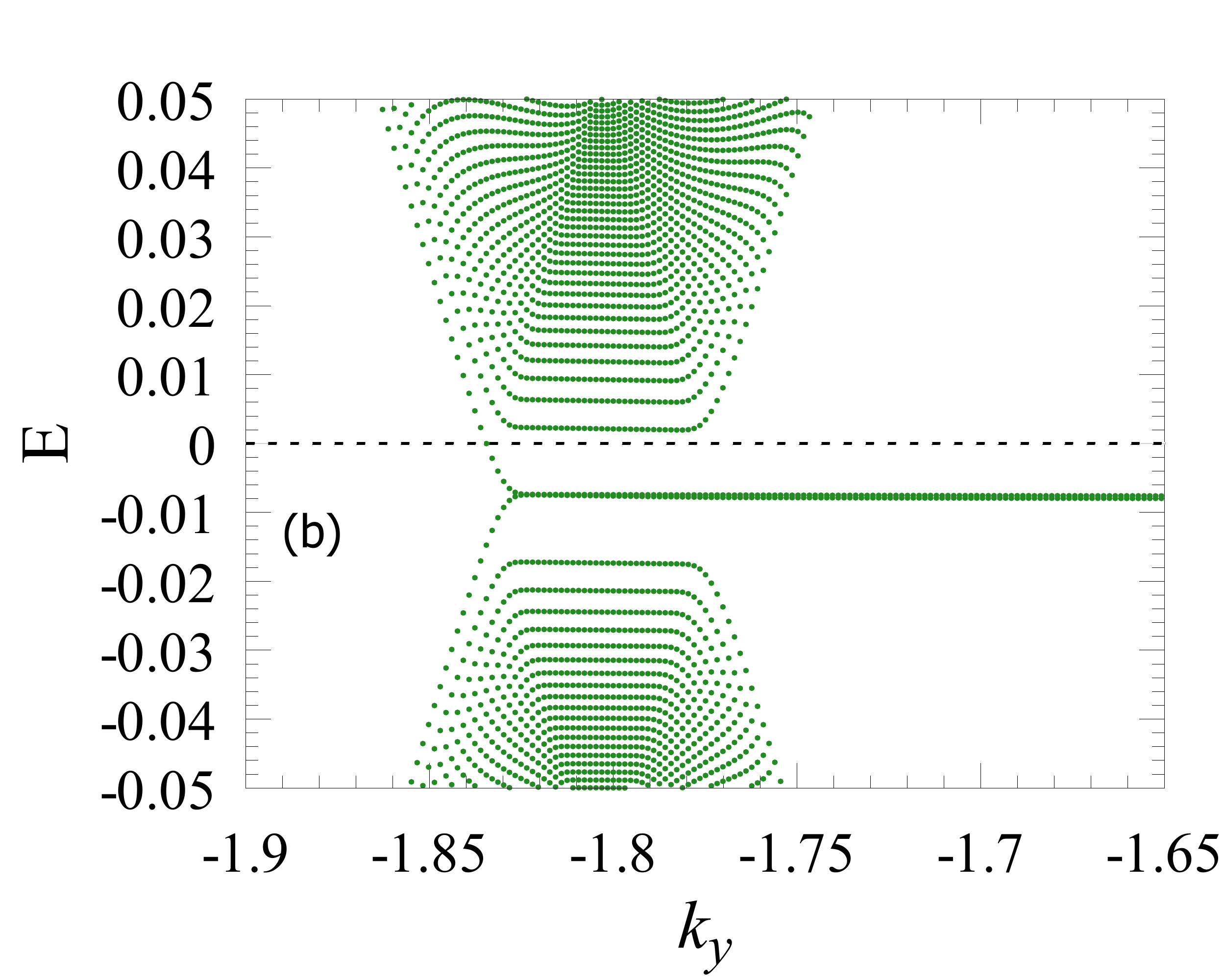}} \quad
 \subfloat{\includegraphics[width=0.6\columnwidth]{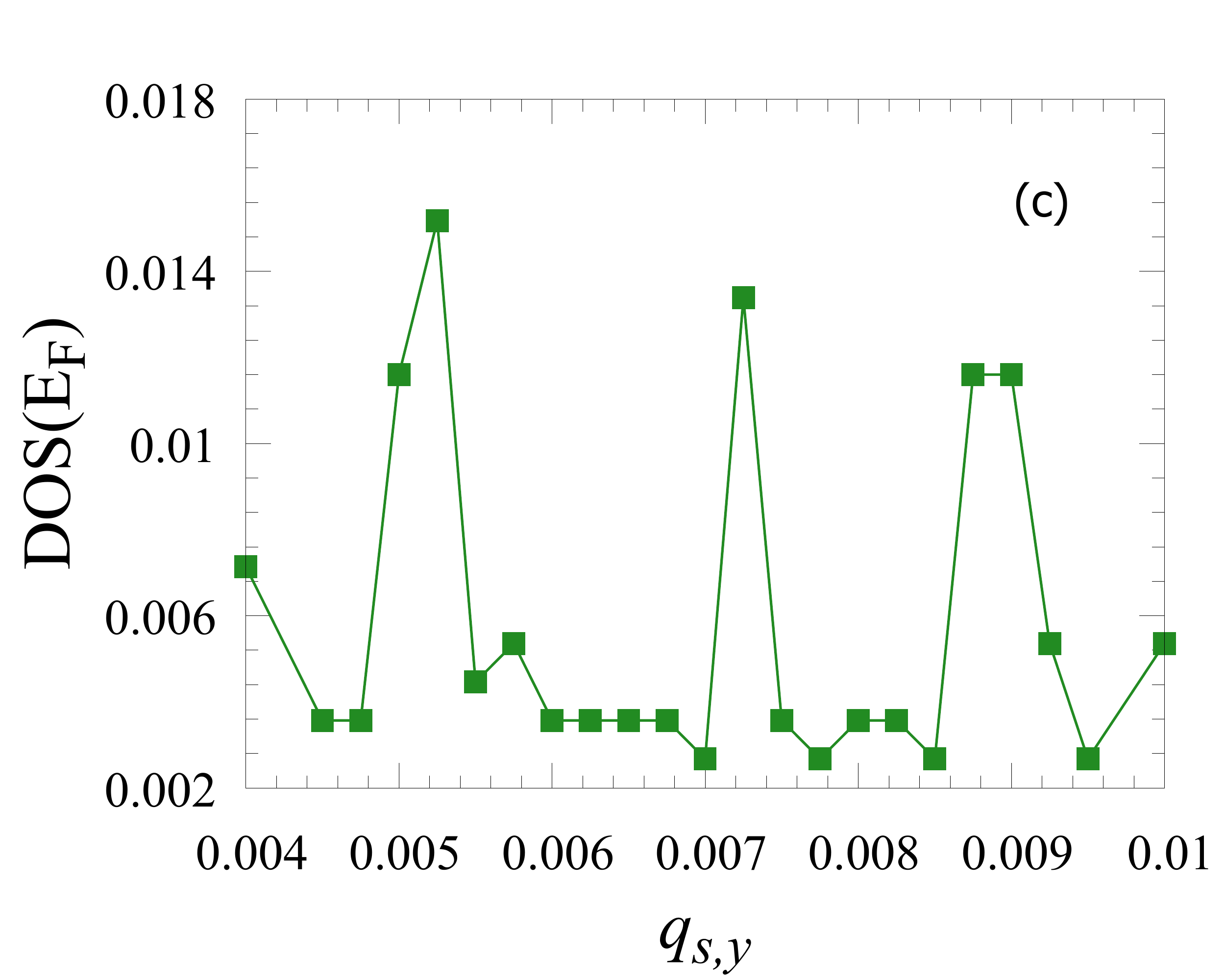}} \\[-3ex]
\subfloat{\includegraphics[width=0.6\columnwidth]{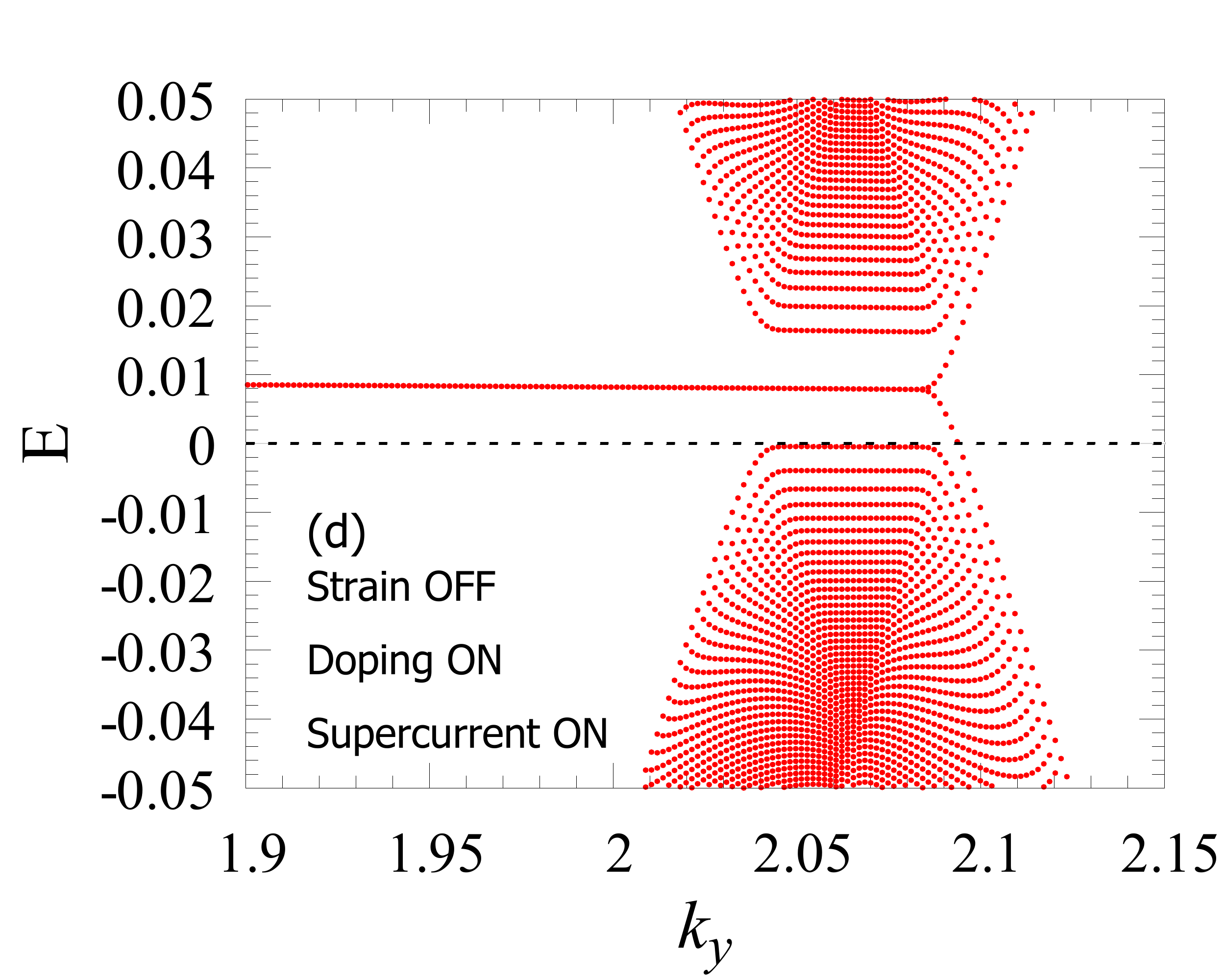}} \quad
 \subfloat{\includegraphics[width=0.6\columnwidth]{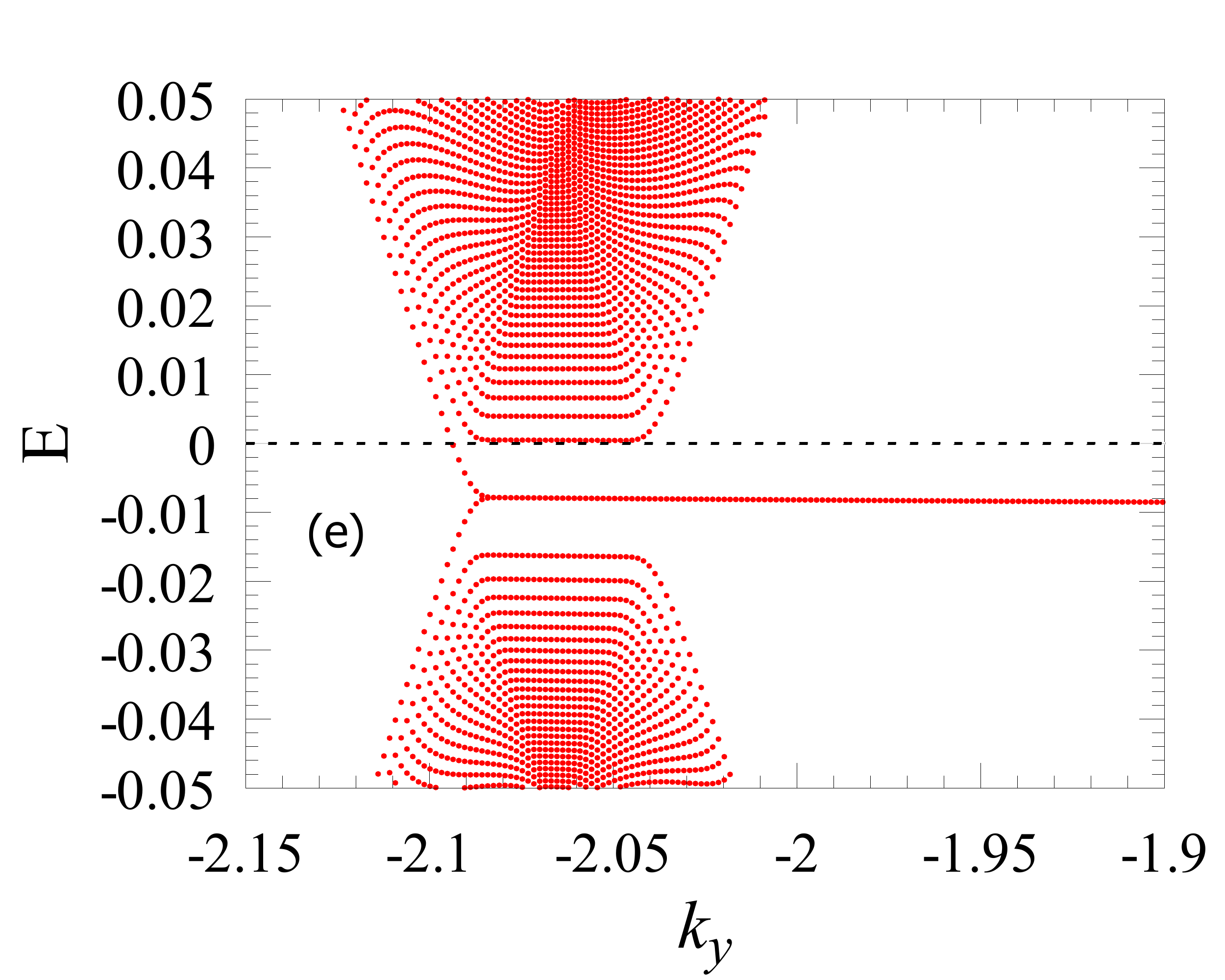}} \quad
\subfloat{\includegraphics[width=0.6\columnwidth]{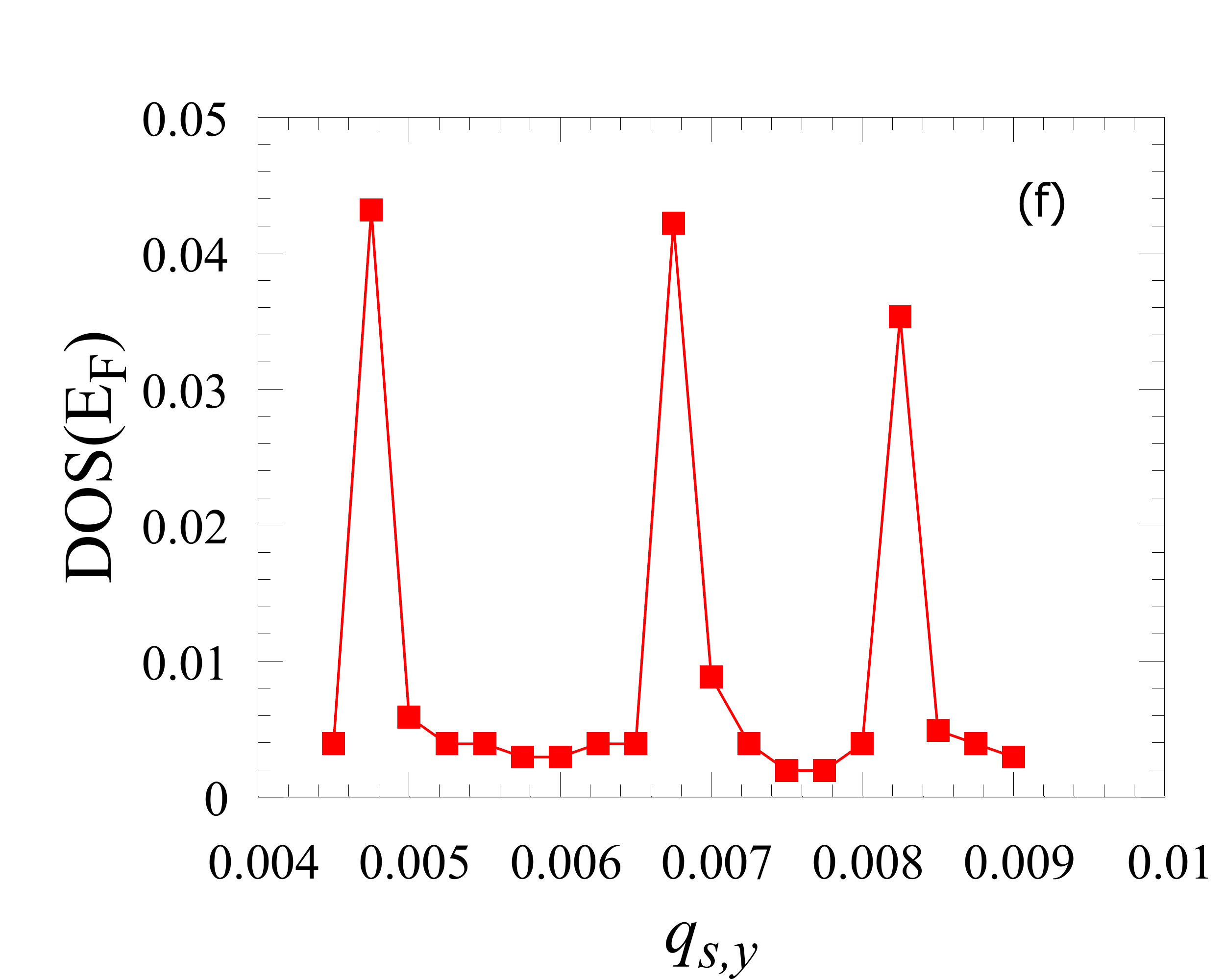}}
\caption{Results of the numerical calculations for 
the 2D $d_{xy}$ SC in the presence of strain and 
doping gradient and of an applied supercurrent. 
(a) The low-energy spectrum around 
node 1 in the presence of strain 
and of an applied supercurrent in the $y$-direction.
(b) Same as (a) for the opposite node 1'. 
Note that the LL's have an opposite shift in energy w.r.t. 
node 1. (c) The DOS at the Fermi level $E_{F}$ as a function of 
the applied supercurrent parameterized by the 
effective momentum $q_{s,y}$~\cite{dG}. 
The sharp spikes 
occur whenever a pair of LL's crosses the Fermi energy. 
(d)-(f) Same as (a)-(c) in the presence of a doping gradient
alone.}
\label{Fig:spr_crnt}
\end{figure*}

\emph{Numerical Results. --}
We model the lattice under strain 
via a slow variation of the hopping 
coefficients on the 
scale of the inter-site spacing. 
For convenience, we consider 
pure shear strain 
as a function of $x$ alone. 
The hopping coefficients 
are modulated as $t \rightarrow t(1\pm l \delta_{sp})$, 
along $x$ and $y$-direction respectively, where $l$ denotes
the position along $x$, and $\delta_{sp}$ is a small 
parameter. In the case of doping gradients 
we allow the chemical potential to vary as 
$\mu \rightarrow \mu(1+ l \delta_{dp})$, 
as a function of the $x$-coordinate alone. 
Consequently, we impose
periodic and open boundary conditions along 
the $y$- and $x$-axes respectively.
In the low-energy limit the modified coefficients 
lead to 
vector potentials (Eqs.~\ref{Eq:Strn_gg},~\ref{Eq:Dpng_grdn}), 
corresponding to 
uniform pseudo-magnetic fields only around the 
$(1, 1')$ pair of nodes.  For
an expanded discussion, please consult the SM.

We find the energy spectra of the modified lattice Hamiltonians
numerically. 
All results are reported 
in units where $ta=1$. The largest
change in either hopping or chemical 
potential over the entire extent of the lattice is
on the order of $10\%$.

In Fig.~\ref{Fig:Zr_crnt} (a) we show the low-energy
spectrum as a function of $k_{y}$ for the 
unperturbed $d_{xy}$ superconductor around the 
node 1. 
Note that the presence of the flat zero-energy 
dispersion is  
associated with topologically-protected 
Majorana edge states~\cite{Potter_PRL_2014} and is unrelated 
to the effect under discussion. In 
the presence of strain, the low-energy spectrum 
around the $(1, 1')$ pair of nodes is re-organized into
discrete, flat bands as illustrated in Fig.~\ref{Fig:Zr_crnt} (b). 
The inter-level energy differences are those
predicted by the Dirac-Landau spectrum.
Similar effects are observed upon 
the inclusion of a weakly-varying 
chemical potential as shown in Fig.~\ref{Fig:Zr_crnt} (c). 
The associated 
total DOS per unit area 
in the low-energy sectors, Figs.~\ref{Fig:Zr_crnt} (d)-(f),
reflects the emergence of LL's.

In the presence of both strain and weak, applied supercurrents
the spectrum around the opposite $(1,1')$ points is shifted to 
higher and lower energies respectively, as shown in Figs.~\ref{Fig:spr_crnt} (a) and 
(b). An analogous behavior is obtained in the case of 
a doping gradient, as shown in Figs.~\ref{Fig:spr_crnt} (d) and (e).
The total DOS per unit area in the presence of an 
applied supercurrent reflects the shifts in energies of the LL's.
Most notably, it exhibits sharp enhancements 
whenever pairs of LL's around opposite nodes cross 
the Fermi energy. This is shown in Figs.~\ref{Fig:spr_crnt} (c) and (f)
for a range of supercurrents for strain and doping gradient
cases, respectively. Note that the applied supercurrent also lifts the degeneracy of
the LL's and leads to a broadening of the sharp peaks in the DOS. 

\emph{Summary and outlook. --}
We proposed that weak, slowly-varying 
strains and doping gradients generically 
give rise to pseudo-magnetic fields in the low-energy limit of
two-dimensional nodal superconductors. 
For simplicity, our discussion was focused 
on a prototypical 2D $d_{xy}$ SC. This is not 
essential 
and similar 
effects are expected to occur in other types of SC which
exhibit Dirac-like spectra around points in the Brillouin Zone. 
Examples include $d_{x^2-y^2}$ cuprates and odd-parity $p$-wave  
cases. In view of similar proposals for pseudo-magnetic fields 
in Weyl semi-metals~\cite{Shapourian,Cortijo,Fujimoto,Pikulin,Grushin,Liu_PRB_2017, Liu_unpb},
such effects are also likely to occur in more complicated 3D systems. 

The most likely candidates for the 
experimental observation of such effects are high-$T_c$
cuprates. Specifically, in 
YBa$_2$Cu$_3$O$_x$ thin films and La$_{2−-x}$Sr$_x$CuO$_4$–-La$_2$CuO$_4$
bilayer samples controlled doping gradients have recently been
achieved \cite{Taylor2015,Wu2013}.  According to our
theory such samples should already exhibit Landau level quantization
which is in principle observable as a series of sharp peaks in DOS through standard
quasiparticle spectroscopies such as the angle-resolved photoemission
(ARPES) or scanning tunneling spectroscopy (STS).  In the above
samples a $\sim10$\% doping change is imposed over a millimeter scale which
allows us  to estimate (SM)  the effective pseudo-magnetic field as
$\mathscr{B}\simeq 0.32$ mT and the
Landau level spacing $\hbar\omega_{c}\simeq 82 \ \mu$eV. While
this is probably too small  to resolve by STS or ARPES we see no
fundamental reason why similar doping variations could not be
imposed  on a 
$\mu$m scale which would produce an effective field of order one Tesla
and clearly observable Landau levels.
Under the applied supercurrent such 
samples would in addition show quantum oscillations in transport properties,
such as the longitudinal thermal conductivity $\kappa_{xx}$. 
We speculate that they might also  exhibit exactly quantized
thermal Hall conductance $\kappa_{xy}$ when the zero of the energy is
tuned to lie between the bulk Landau levels by supercurrent as in
Fig.\ 3  and the electronic contribution to $\kappa_{xx}$ vanishes. In this situation each bulk band is expected to carry a
nonzero Chern number and produce a protected chiral edge mode, already
visible in Fig.\ 3 as a dispersing feature terminating the flat
Landau levels.  Finally, large local strain gradients have recently been observed in
nanocomposite  YBa$_2$Cu$_3$O$_{7-\delta}$ films \cite{Guzman2017}
  which may afford an opportunity to study the effects discussed in
  this paper on nanoscale, similar to the seminal work on
  ``nanobubbles'' in graphene \cite{Levy_Science_2010}.

During the preparation of this manuscript, we became 
aware of the similar results reported in Ref.~\cite{Massarrelli_arxiv_2017}.

\begin{acknowledgments}
{\it Acknowledgements. --} The authors are indebted to T. Liu for illuminating discussions and  thank NSERC, CIfAR and Max Planck - UBC Centre for Quantum Materials for support.  
\end{acknowledgments}


\pagebreak
\widetext
\begin{center}
\textbf{\large Landau levels from neutral 
Bogoliubov particles in two-dimensional nodal superconductors under 
strain and doping gradients :\\[+2ex] 
Supplementary Material}
\end{center}

\setcounter{equation}{0}
\setcounter{figure}{0}
\setcounter{table}{0}
\setcounter{page}{1}
\makeatletter
\renewcommand{\theequation}{S\arabic{equation}}
\renewcommand{\thefigure}{S\arabic{figure}}
\renewcommand{\bibnumfmt}[1]{[S#1]}
\renewcommand{\citenumfont}[1]{S#1}

\title{Supplementary Material}
\author{Emilian M.\ Nica, Marcel Franz}
\date{\today}

\maketitle

In Sec.~I (a) and (b)
we define the lattice pairing Hamiltonian and derive 
the effect of an arbitrary but slowly-varying deformation in 
the low-energy, continuum limit. In Sec.~I (c)
we give the most general expression for the resulting 
vector potentials. The analogous case of a doping gradient is briefly discussed in Sec. I (d).
In Sec. I (e), we write the explicit form of the 
low-energy Hamiltonian (Eq. 4 of the main text). 
The effective lattice models used in the numerical 
calculations are discussed in Sec. II.
Finally, Sec. III gives the estimates for the LL spacing and 
pseudo-magnetic fields reported in the main text.

\section{I.~Low-energy, continuum limit in the presence 
of arbitrary deformations and doping gradients}

The Hamiltonian on the two-dimensional square lattice is given by 
\enni \enbe
\hat{H}=\hat{H}_{TB} + \hat{H}_{\Delta}
\label{Eq:Hmlt}
\enee
\enni where $\hat{H}_{TB}$ is a nearest-neighbor (NN) tight-binding part, 
and $\hat{H}_{\Delta}$ is a pairing part corresponding to a
$d_{xy}$ irreducible representation of the $D_{4h}$ point group. 

\subsection{a.~Tight-binding part in the 
presence of an arbitrary deformation}

In the absence of strain, the TB part is given by
\enni \enbe
\label{Eq:H_TB}
\hat{H}_{TB}= \sum_{i} \left[ \sum_{j \in \braket{ij}} \sum_{\sigma} 
t(\bm{\delta}_{j}) c^{\dagger}_{\sigma}(\bm{R}_{i}) 
c_{\sigma}(\bm{R}_{i} + \bm{\delta}_{j}) + h.c. \right] 
- \sum_{i} \mu c^{\dagger}_{\sigma}(\bm{R}_{i}) c_{\sigma}(\bm{R}_{i}),
\enee
\enni where $\sigma$ is a spin index which is ignored 
for the rest of this subsection. 
$\bm{R}_{i}$ are the Bravais lattice vectors, 
while $\bm{\delta}_{j} \in \{(a, 0), (0, a) \}$ are the vectors which 
connect NN's, $a$ is the NN distance, and $\mu$ is 
the chemical potential. The latter is chosen s.t. 
the system in below half-filing. 
$\hat{H}_{TB}$ can is diagonalized by applying a 
Fourier transform
\enni \enbe
\label{Eq:FT}
c(\bm{R}_{i})=\frac{1}{\sqrt{N}} \sum_{\bm{k} \in BZ} e^{i \bm{k} \cdot \bm{R}_{i}} c_{\bm{k}},
\enee
\enni where $N$ is the number of unit cells. 

In the presence of a lattice deformation, 
the lattice TB Hamiltonian undergoes the transformation
\enni \enbe
\label{Eq:Trns_lttc}
\bm{R}_{i} \rightarrow \bm{R}^{'}_{i} = \bm{R}_{i} + \bm{\epsilon}(\bm{R}_{i}),
\enee
\enni where $\bm{\epsilon}(\bm{R}_{i})$ is a position-dependent displacement. 
In general, the hopping coefficients are modified accordingly:
\enni \begin{align}
\label{Eq:Hppn_cffc}
t(\bm{\delta}_{j}) \rightarrow t'(\bm{\delta}_{j}) = & t\left[ \bm{R}_{i} +  \bm{\delta}_{j} +  \bm{\epsilon}(\bm{R}_{i}  + \bm{\delta}_{j}) 
- \bm{R}_{i} - \bm{\epsilon}(\bm{R}_{i}) \right] \notag \\
= &  t \left[ \bm{\delta}_{j} + \bm{\epsilon}(\bm{R}_{i} + \bm{\delta}_{j}) - \bm{\epsilon}(\bm{R}_{i}) \right].
\end{align}
We assume that the 
hopping coefficients can be approximated by 
continuous functions of the displacement.
The transformed Hamiltonian is
\enni \begin{align}
\hat{H}_{TB} \rightarrow \hat{H}^{'}_{TB} = & \sum_{i} \left[ \sum_{j \in \braket{ij}} t \left[\bm{\delta}_{j} + \bm{\epsilon}(\bm{R}_{i} + \bm{\delta}_{j}) - \bm{\epsilon}(\bm{R}_{i}) \right] c^{\dagger}(\bm{R}_{i} + \bm{\epsilon}(\bm{R}_{i})) c(\bm{R}_{i} + \bm{\epsilon}(\bm{R}_{i}+  \bm{\delta}_{j} ) + \bm{\delta}_{j}) + h.c. \right] \notag \\
& - \sum_{i} \mu c^{\dagger}(\bm{R}_{i} + \bm{\epsilon}(\bm{R}_{i}))c(\bm{R}_{i} + \bm{\epsilon}(\bm{R}_{i})).
\end{align}

We consider the low-energy, continuum limit 
of $\hat{H}^{'}_{TB}$. Consequently, we approximate the lattice 
operators as products of parts which vary rapidly and slowly on the scale of the lattice as
\enni \enbe
c(\bm{R}_{i}) \approx \sum_{\alpha} e^{i \bm{K}_{\alpha} \cdot \bm{R}_{i}} \Psi^{(\alpha)}(\bm{R}_{i}),
\enee
\enni where $\Psi^{(\alpha)}(\bm{R}_{i})$ is a slowly-varying
Fermi field and $\alpha \in \{1, 1', 2, 2'\}$ represents the 
positions of the four nodes at Fermi wave-vectors $\bm{K}_{\alpha} \in \{(0, \pm K_{F}), (\pm, K_{F}, 0) \}$.
We assume that the lattice displacements can be approximated 
by a continuous displacement field:
\enni \begin{align}
\bm{R}_{i} + \epsilon(\bm{R}_{i}) \rightarrow \bm{r} + \bm{\epsilon}(\bm{r}).
\label{Eq:Dfrm}
\end{align}
\enni On the scale of the lattice, variations in the displacements can be approximated by
\enni \begin{align}
\bm{\epsilon}(\bm{r} + \bm{\delta}_{j}) \approx \epsilon(\bm{r}) + (\bm{\delta}_{j} \cdot \bm{\nabla}) \bm{\epsilon}(\bm{r}).
\label{Eq:dfrm_expn}
\end{align}
\enni In the following, we expand the continuum limit 
of the Hamiltonian in terms of leading gradient terms. 

It should be noted that such an expansion is valid provided that
the gradient term in Eq.~\ref{Eq:dfrm_expn} remains small throughout. 
As discussed in in the following, in order to obtain finite pseudo-magnetic fields
we consider deformation gradients which vary monotonically. 
This implicitly introduces a spatial scale at which these gradients are no longer 
small.
Therefore, the effective continuum approximation only holds 
provided that the deformation at any point of a sample of finite extent remains
small. 
 
Applying the above to $\hat{H}^{'}_{TB}$, we obtain
\enni \begin{align}
\hat{H}^{'}_{TB} = & \int d^{2} r \sum_{\alpha} \Psi^{\dagger, (\alpha)}(\bm{r}+\bm{\epsilon}) 
\Bigg\{ \sum_{j} 2 t \bigg( \bm{\delta}_{j} + (\bm{\delta}_{j} \cdot \bm{\nabla}) \bm{\epsilon} \bigg) 
\bigg[ \cos \bigg( \bm{K}_{\alpha} \cdot  (\bm{\delta}_{j} + (\bm{\delta}_{j} \cdot \bm{\nabla}) \bm{\epsilon}) \bigg) \notag \\
+ & i \sin\bigg( \bm{K}_{\alpha} \cdot  (\bm{\delta}_{j} + (\bm{\delta}_{j} \cdot \bm{\nabla}) \bm{\epsilon}) \bigg)  (\bm{\delta}_{j} \cdot \bm{\nabla}) \bigg] \Bigg\} \Psi^{(\alpha)}(\bm{r}+\bm{\epsilon}) 
-  \mu  \sum_{\alpha} \int d^{2} r 
\Psi^{\dagger, (\alpha)}(\bm{r}+\bm{\epsilon}) \Psi^{(\alpha)}(\bm{r}+\bm{\epsilon}),
\end{align}
\enni where we neglected inter-node terms. The explicit dependence of the fields 
on $\bm{\epsilon}$ can be formally eliminated via a coordinate transformation
\enni \enbe
\bm{r}^{'} = \bm{r} + \bm{\epsilon}(\bm{r}),
\enee
\enni with a Jacobian $1+\epsilon_{ii}$, where $\epsilon_{ij} =\partial_{j} \epsilon_{i}$
and implicit summation is assumed. We also 
approximate $\bm{\delta}_{j} \cdot \bm{\nabla} \epsilon(\bm{r}) \approx \bm{\delta}_{j} \cdot \bm{\nabla} \epsilon(\bm{r}^{'})$
and expand the following terms
\enni \begin{align}
 t \bigg( \bm{\delta}_{j} + (\bm{\delta}_{j} \cdot \bm{\nabla}) \bm{\epsilon} \bigg) \approx &
t(\bm{\delta}_{j}) + (\bm{\delta}_{j} \cdot \bm{\nabla}) \bm{\epsilon} \cdot \bm{\nabla} t(\delta_{j}), \\
\cos \bigg( \bm{K}_{\alpha} \cdot  (\bm{\delta}_{j} + (\bm{\delta}_{j} \cdot \bm{\nabla}) \bm{\epsilon}) \bigg) \approx &
\cos ( \bm{K}_{\alpha} \cdot  \bm{\delta}_{j}) - \bm{K}_{\alpha} \cdot (\bm{\delta}_{j} \cdot \bm{\nabla}) \bm{\epsilon}
\sin\left( \bm{K}_{\alpha} \cdot \bm{\delta}_{j} \right), \\
\sin \bigg( \bm{K}_{\alpha} \cdot  (\bm{\delta}_{j} + (\bm{\delta}_{j} \cdot \bm{\nabla}) \bm{\epsilon}) \bigg) \approx &
\sin ( \bm{K}_{\alpha} \cdot  \bm{\delta}_{j}) + \bm{K}_{\alpha} \cdot (\bm{\delta}_{j} \cdot \bm{\nabla}) \bm{\epsilon}
\cos\left( \bm{K}_{\alpha} \cdot \bm{\delta}_{j} \right).
\end{align}
\enni To zeroth order in the gradient expansion we obtain
\enni \begin{align}
\label{Eq:Hmlt_TB_0}
 H^{(0)}_{TB}= & \int d^{2} r  \sum_{\alpha} \sum_{j} 2 t ( \bm{\delta}_{j} ) 
\Psi^{\dagger, (\alpha)}(\bm{r}) 
 \Bigg\{ \cos{\left(\bm{K}_{\alpha} \cdot \bm{\delta}_{j} \right) }
+ i \sin \left( \bm{K}_{\alpha} \cdot  \bm{\delta}_{j} \right)
\left(\bm{\delta}_{j} \cdot \bm{\nabla} \right)  \Bigg\} \Psi^{(\alpha)}(\bm{r})
 \notag \\
-  & \mu  \sum_{\alpha} \int d^{2}r
\Psi^{\dagger, (\alpha)}(\bm{r}) \Psi^{(\alpha)}(\bm{r}).
\end{align}
\enni The first and last terms cancel, 
since $\bm{K}_{\alpha}$ is on the Fermi surface. 
The second term gives the leading linear 
dispersion and also defines the Fermi velocity 
$v_{F}= 2t a \sin(K_{F}a)$.

To first order, we obtain
\enni \begin{align}
\label{Eq:Gnrl_TB:Hmlt}
H^{(1)}_{TB} = & 
\sum_{\alpha}  \sum_{j}  \Bigg\{  \int d^{2} r( \epsilon_{xx} + \epsilon_{yy}) 2 t (\bm{\delta}_{j})
\cos \left(\bm{K}_{\alpha} \cdot  \bm{\delta}_{j}  \right) 
+ \int d^{2} r 2 (\bm{\delta}_{j} \cdot \bm{\nabla}) \bm{\epsilon} \cdot \bm{\bm{\nabla}}t
\cos \left( \bm{K}_{\alpha} \cdot  \bm{\delta}_{j} \right) \notag \\
& + \int d^{2} r 2 t (\bm{\delta}_{j})
\left( -  \bm{K}_{\alpha} \cdot (\bm{\delta}_{j} \cdot \bm{\nabla}) \bm{\epsilon}
\sin\left( \bm{K}_{\alpha} \cdot \bm{\delta}_{j} \right)  \right) 
\Bigg\} 
\Psi^{\dagger, (\alpha)}(\bm{r}) 
 \Psi^{(\alpha)}(\bm{r}) \notag \\
-  & \mu  \sum_{\alpha} \int d^{2}r (\epsilon_{xx} + \epsilon_{yy})
\Psi^{\dagger, (\alpha)}(\bm{r}) 
 \Psi^{(\alpha)}(\bm{r}).
\end{align}
\enni The first two terms cancel independently 
of the details of the deformation. These  
represent a local dilation/contraction
with all other parameters fixed. The 
expression can be simplified further by 
carrying out the summations over NN's. The 
result is summarized by
\enni \begin{align}
H^{(1)}_{TB} = & 
\sum_{\alpha}  \int d^{2} r e \bm{\mathscr{A}}^{\alpha}(\bm{r})
\Psi^{\dagger, (\alpha)}(\bm{r}) 
 \Psi^{(\alpha)}(\bm{r}),
\end{align}
\enni where the electron charge $e$ was introduced for 
dimensional consistency. 
The effective \emph{vector} potentials around
either $(1, 1')$ nodes are defined as
\enni \begin{align}
\label{Eq:d_xy_g_fld_y}
\bm{\mathscr{A}}^{(1, 1')} =
\begin{pmatrix}
0 \\
 \left( \frac{2a}{e} \right) \bigg[t'_{\parallel}  \epsilon_{xx}
+ t'_{\perp} \epsilon_{yx}
+  t'_{\perp}  \cos \left(K_{F} a  \right) \epsilon_{xy} 
+  \bigg( t'_{\parallel} \cos \left(K_{F} a  \right) 
-  t K_{F} \sin \left(K_{F} a \right) \bigg) \epsilon_{yy} \bigg] 
\end{pmatrix}
\end{align}
\enni The $\mathscr{A}_{x}$ component
is formally set to zero since there is no linear-derivative term along 
the $x$-direction to leading order for the TB part. We also defined the 
coefficients $\partial_{x}t_{x} = \partial_{y}t_{y}=  t'_{\parallel}, 
\partial_{x}t_{y} = \partial_{y}t_{x}= t'_{\perp}$, 
which are restricted by the symmetry of the square lattice, 
while, in general, $ t'_{\perp}  \neq  t'_{\parallel}$. 
The analogous non-trivial vector potentials $\mathscr{A}_{x}$ around the other
pair of Fermi momenta can be obtained by replacing $x \leftrightarrow y$. 

\subsection{b.~Pairing part in the 
presence of an arbitrary deformation}

In the absence of any deformation, the paring part of the Hamiltonian 
for a 2D $d_{xy}$ SC is given by
\enni \begin{align}
\hat{H}_{\Delta} = \sum_{i} \sum_{j \in \braket{\braket{ij}}} \sum_{\sigma} \sum_{\sigma'}
\Delta_{\sigma \sigma'}(\bm{\delta}_{j}) \left[ c_{\sigma}(\bm{R}_{i}) c_{\sigma'}(\bm{R}_{i} + \bm{\delta}_{j}) +  
c_{\sigma}(\bm{R}_{i}) c_{\sigma'}(\bm{R}_{i} - \bm{\delta}_{j}) \right] + h.c. , 
\end{align}
\enni where the pairing occurs for next-nearest neighbor (NNN) sites, 
$\Delta_{\sigma \sigma'}(\bm{\delta}_{j}) = \Delta(\bm{\delta}_{j}) i \sigma_{y}$ 
corresponding to even-parity, spin-singlet pairing. In addition, 
$\Delta(\bm{\delta}_{1/2}) = \pm \Delta$, with 
$\bm{\delta}_{1/2}=(\pm a,a)$ determining the vectors 
which connect NNN's.

We allow for the deformation given 
by Eq.~\ref{Eq:Dfrm} and assume 
that it's effect on the pairing potential
can be generically captured to lowest order in the strains, 
The derivation of the low-energy, continuum limit is 
analogous to that of the TB part. We list the final results 
for the $(1, 1')$ pair of nodes:
\begin{align}
H^{(0)}_{\Delta}= & \int d^{2}r \sum_{(\alpha, \beta)}
 4 a  \Delta_{\sigma \sigma'}
\sin\left( K_{Fy, \alpha} a  \right)
\Psi^{\alpha}_{\sigma}(\bm{r}) \left( - i \partial_{x} \right) \Psi^{\beta}_{\sigma'}(\bm{r})  
+ h.c. , 
\label{Eq:Hmlt_prng_0}
\end{align}
\enni \begin{align}
\label{Eq:d_xy_strn_crrc}
H^{(1)}_{\Delta}
= &  \int d^{2}r \sum_{(\alpha, \beta)} \sum_{\sigma, \sigma'}  (-i\sigma_{y})_{\sigma, \sigma'}
 4a \bigg [\left( \epsilon_{xx} \partial_{x} \Delta + \epsilon_{yx} \partial_{y} \Delta \right)
\cos\left( K_{F} a \right)  
- \Delta K_{F} \epsilon_{yx} \sin\left(K_{F} a \right)
\bigg]
\Psi^{\alpha}_{\sigma}(\bm{r}) \Psi^{\beta}_{\sigma'}(\bm{r}),
\end{align}
\enni where $\alpha \neq \beta$. The zeroth-order term 
defines a velocity $v_{\Delta}=4a \Delta \sin(K_{F}a)$.
For the $(2,2')$ pair, we replace $x \leftrightarrow y$. 
We can also set $\partial_{x} \Delta=\partial_{y} \Delta=\Delta^{'}$.

\subsection{c.~Effective gauge potentials in the presence of strain}

The first-order corrections to the pairing part 
can be eliminated via the gauge transformation
\enni \begin{align}
\Psi^{\beta}_{\sigma'} \rightarrow \Psi^{\beta}_{\sigma'} e^{-i \text{sgn}(K_{Fy, \alpha}) \frac{ \phi(\bm{r}) }{v_{\Delta}}},
\end{align}
\enni where 
\enni \begin{align}
\phi = 4a \Delta^{'} \left( \epsilon_{x} + \epsilon_{y} \right) \cos\left( K_{F, \alpha} a \right) 
- \Delta K_{F} \epsilon_{y} \sin\left(K_{F} a \right).
\end{align}
\enni The transformation modifies $\mathscr{A}_{y}$ in Eq.~\ref{Eq:d_xy_g_fld_y}
to 
\enni \enbe
\mathscr{A}^{'}_{y} = \mathscr{A}_{y} - \frac{v_{F}}{v_{\Delta}e} \partial_{y} \phi.
\enee
\enni The terms proportional 
to $\sin(K_{F}a)$ cancel.
The expression for the most general gauge strain-induced 
vector potentials for the $(1, 1')$ fields reduces to
\enni \begin{align}
\mathscr{A}_{y} 
= & \left( \frac{2a}{e} \right) \Bigg\{ t'_{\parallel}  \epsilon_{xx}
+  \bigg[ t'_{\parallel} - \left( \frac{t}{\Delta} \right) \Delta^{'} \bigg] \cos \left(K_{F} a  \right)  \epsilon_{yy}
+ t'_{\perp} \epsilon_{yx}
+  \bigg[ t'_{\perp} - \left( \frac{t}{\Delta} \right) \Delta^{'}  \bigg] \cos \left(K_{F} a  \right) \epsilon_{xy} \Bigg\}.
\end{align}
\enni The corresponding $\mathscr{A}_{x}$ for the $(2,2')$ pair is obtained 
by replacing $x \leftrightarrow y$. Note that 
$t'_{\parallel}$ corresponds to a change in bond length, 
$t'_{\perp}$ to a change change in bond angle, 
while $\Delta^{'}$ includes both. In the most general case, 
the coefficients depend on the details of the model, 
and in particular on the symmetry of the orbitals. 

We focus on cases where the dominant contribution 
comes from the change in bond length i.e.
$t'_{\parallel} \neq 0, t'_{\perp} \approx 0$.
Such an approximation can be justified in principle 
using a Slater-Koster scheme~\cite{Slater_PR_1954}, 
and is consistent with the similar 
case of graphene~\cite{Castro_Neto_RMP_2009}.
Additionally, we assume that the leading change in the pairing potential 
under strain $\Delta$ is also negligible. 
Under these assumptions, and writing
\enni \begin{align}
a  t'_{\parallel} = t \frac{d \ln t}{d \ln a}
= t \beta,
\end{align}
\enni we obtain the form of the vector potentials
discussed in the main text.

\subsection{d.~Effective gauge potentials from doping gradients}

We consider the effect of a doping gradient $\mu \rightarrow \mu(\bm{R}_{i})$
on the Hamiltonian of Eq.~\ref{Eq:Hmlt}. 
The low-energy, continuum limit in this case is
\enni \begin{align}
H= H^{(0)}_{TB} + H^{(0)}_{\Delta}+ H_{dg},
\end{align}
\enni where the first two terms correspond to the unperturbed 
Hamiltonians for the TB and pairing part, given by Eqs.~\ref{Eq:Hmlt_TB_0} and
~\ref{Eq:Hmlt_prng_0}, respectively. The last term 
is the contribution of a spatially-varying chemical potential
\enni \begin{align}
H_{dg} = -  & e \sum_{\alpha} \int d^{2}r \left(\frac{\tilde{\mu}(\bm{r})}{e} \right)
\Psi^{\dagger, (\alpha)}(\bm{r}) \Psi^{(\alpha)}(\bm{r}),
\end{align}
\enni where
\enni \begin{align}
\tilde{\mu}(\bm{r}) = \mu g(\bm{r}).
\end{align}
The corresponding vector potential around the $(1, 1')$
pair of nodes is
\enni \begin{align}
\bm{\mathscr{A}}^{(1, 1')} =
\begin{pmatrix}
0 \\
\left( \frac{\mu}{e} \right) g(\bm{r})
\end{pmatrix},
\end{align}
\enni with an analogous $\mathscr{A}_{x}$ around $(2, 2')$.
These vector potentials must remain small throughout 
the finite area of the sample and are subject to the constraints 
imposed in the case of strain. 

\subsection{e. Low-energy Nambu form of the Hamiltonian in the presence of non-trivial vector potentials}

In either strain or doping cases, we can write 
the Hamiltonian in the low-energy sector as
\enni \begin{align}
H= \int d^{2} r
\begin{pmatrix}
\Psi^{\dagger, (1)}_{\uparrow} \\
\Psi^{\dagger, (1')}_{\downarrow} \\
\Psi^{(1)}_{\uparrow} \\
\Psi^{(1')}_{\downarrow} 
\end{pmatrix}^{T} 
\begin{pmatrix}
v_{F} \bigg( i \partial_{y} + e \frac{\mathscr{A}_{y}}{v_{F}} \bigg) & 0 & 0 & v_{\Delta} (-i \partial_{x}) \\
 0  & v_{F} \bigg( - i \partial_{y} + e \frac{\mathscr{A}_{y}}{v_{F}} \bigg) & v_{\Delta} (-i \partial_{x}) & 0 \\
 0  & v_{\Delta} (-i \partial_{x}) &v_{F} \bigg( i \partial_{y} - e \frac{\mathscr{A}_{y}}{v_{F}} \bigg) & 0 \\
v_{\Delta} (-i \partial_{x}) & 0 & 0 &v_{F} \bigg( - i \partial_{y} - e \frac{\mathscr{A}_{y}}{v_{F}} \bigg) 
\end{pmatrix}
\begin{pmatrix}
\Psi^{(1)}_{\uparrow} \\
\Psi^{(1')}_{\downarrow} \\
\Psi^{\dagger, (1)}_{\uparrow} \\
\Psi^{\dagger, (1')}_{\downarrow}
\end{pmatrix},
\end{align}
\enni which is identical to Eq.~4 of the main text. 
With vanishing vector potentials, one can easily check 
that this is the low-energy, continuum limit of the 
Hamiltonian of Eq. 1 main text:
\enni \begin{align}
H= \sum_{\bm{k}}
\begin{pmatrix}
c^{\dagger}_{\bm{k} \uparrow} \\
c_{- \bm{k} \downarrow}
\end{pmatrix}^{T} 
\begin{pmatrix}
h_{\bm{k}} & \Delta_{\bm{k}} \\
\Delta_{\bm{k}} & - h_{\bm{k}}
\end{pmatrix}
\begin{pmatrix}
c_{\bm{k} \uparrow} \\
c^{\dagger}_{- \bm{k} \downarrow}
\end{pmatrix}
+ 
\begin{pmatrix}
c^{\dagger}_{\bm{k} \downarrow} \\
c_{- \bm{k} \uparrow}
\end{pmatrix}^{T} 
\begin{pmatrix}
h_{\bm{k}} & - \Delta_{\bm{k}} \\
- \Delta_{\bm{k}} & - h_{\bm{k}}
\end{pmatrix}
\begin{pmatrix}
c_{\bm{k} \downarrow} \\
c^{\dagger}_{- \bm{k} \uparrow}
\end{pmatrix}.
\end{align}

The BdG equations can be obtained in standard fashion 
by considering the block Hamiltonian 
\begin{align}
H^{'} = \int d^{2} r
\begin{pmatrix}
\Psi^{\dagger, (1')}_{\downarrow} \\
\Psi^{(1)}_{\uparrow}
\end{pmatrix}^{T} 
\begin{pmatrix}
v_{F} \bigg( -i\partial_{y} + e \frac{\mathscr{A}_{y}}{v_{F}} \bigg) & v_{\Delta} (-i \partial_{x}) \\
 v_{\Delta} (i \partial_{x} ) & v_{F} \bigg(i\partial_{y} - e \frac{\mathscr{A}_{y}}{v_{F}} \bigg) 
\end{pmatrix}
\begin{pmatrix}
\Psi^{(1')}_{\downarrow} \\
\Psi^{\dagger, (1)}_{\uparrow},
\end{pmatrix}
\end{align}
\enni together with the fact that $\Psi^{(1')}_{\downarrow}, \Psi^{(1)}_{\uparrow}$ are
related through time-reversal. The remaining sector can 
also be obtained via the same operation. 

\section{II.~Numerical calculations}

For the purpose of numerical computation, 
we introduce position-dependent tight-binding coefficients or
chemical potentials at the level of the Hamiltonian
in Eq.~\ref{Eq:Hmlt}. Moreover,
these vary slowly on the scale of the lattice. 
Depending on the spatial dependence of these parameters, 
we recover the different cases discussed in the main text. 

In practice, we allow a non-trivial 
spatial variation along the $x$-direction, 
but keep periodic boundary conditions along 
$y$. 

\subsection{a.~Deformation-induced vector potentials}

In this case, we choose
\enni \begin{align}
t(\bm{\delta}_{j}) \rightarrow t(\bm{\delta}_{j})\bigg[ 1+ f(\bm{R}_{i}, \bm{\delta_{j}}) \bigg].
\end{align}
\enni In the low-energy continuum approximation,
the corrections to the hopping coefficients are 
analogous to the second term in Eq.~\ref{Eq:Gnrl_TB:Hmlt} which is the 
contribution of the transformed TB coefficients:
\enni \begin{align}
H^{(1)}_{TB} = & 
\sum_{\alpha}  \sum_{j}  \int d^{2} r 2 t f(\bm{r}, \bm{\delta}_{j})
\cos \left( \bm{K}_{\alpha} \cdot  \bm{\delta}_{j} \right) 
\Psi^{\dagger, (\alpha)}(\bm{r}) \Psi^{(\alpha)}(\bm{r}) .
\end{align}

We consider the three limiting cases discussed
in the main text:

(i) \emph{Uniaxial strain along $x$ axis}. 
This corresponds to $f(x, \bm{\delta}_{x}) \neq 0$, with 
the component in the $y$ direction equal to 0. 
The vector potential is $\mathscr{A}_{y}= t f(x, \bm{\delta}_{x})$
for the $(1, 1')$ pair of nodes, and $\mathscr{A}_{y}=0$ for 
the other pair. 
On the lattice this amounts to $f(x_{l}, \bm{\delta}_{x})= l \delta_{sp}$, 
where $\delta_{sp} \ll 1$.

(ii) \emph{Hydrostatic compression/dilatation.}
In this case, the finite vector potentials 
about each pair of nodes are equal. 
We take $f(x,\bm{\delta}_{x})=f(x,\bm{\delta}_{y})=f(x)$.
The dependence on $x$ alone is purely for computational convenience.
The resulting vector potentials are $\mathscr{A}_{x/y}= 2t f(x) \big(1+\cos( K_{F} a)\big)$. 
On the lattice, we have $f(x_{l}, \bm{\delta}_{x})= l \delta_{sp}$.

(iii) \emph{Pure shear deformation.}
We choose $f(x,\bm{\delta}_{x})=- f(x,\bm{\delta}_{y})$.
The vector potentials are $\mathscr{A}_{y/x} = \pm 2t f(x)\big(1 - \cos( K_{F} a)\big)$. 
On the lattice, we have $f(x_{l}, \bm{\delta}_{x})= l \delta_{sp}$,
with a corresponding negative sign for hopping along the $y$ direction.   

\subsection{b.~Doping gradient-induced vector potentials}

This case is qualitatively similar to that of a deformation, since
$\tilde{\mu}(\bm{r}) = \mu g(\bm{r})$. In practice, 
we take $g(\bm{r})=g(x)$. On the lattice, this amounts to
$g(x_{l})=l \delta_{dp} $, with $\delta_{dp} \ll 1$.   

\section{III.~Estimate of doping gradient-induced Landau level spacing and pseudo-magnetic fields}

We estimate that the candidate cuprate films are generically 
characterized by the parameters
\\

1) $\Delta \approx $ 30 meV~\cite{Hashimoto_Nat_Phys_2014}
\\

2) $a \approx 3.9$ \AA .
\\

3) $K_{F} a \approx \pi/2$.
\\

4) $t \approx 0.38$  eV~\cite{Korshunov_Phys_C_2004}.
\\

5) $\mu \approx 1.52$ eV
as determined from  $\mu \approx 2t(1 + \cos(K_{F}a) \approx 4t$.
\\

In addition, we estimate $\partial_{x}g(x)\approx $  0.047 mm$^{-1}$ from the 
relative variation in hole concentration $\partial_{x} g(x) \approx (\Delta p)/(p_{0} \Delta x)$ 
over sample length in Ref.~\onlinecite{Taylor_PRB_2015}. The rate 
in Ref.~\onlinecite{Wu_Nat_Mater_2013} is roughly half of this. 
Upon including factors of $\hbar$ for dimensional consistency we 
obtain
\enni \begin{align}
E_{c, \text{Doping}} = & \sqrt{8 \Delta \mu a (\sin(K_{F}a))\partial_{x} g(x)} \\
\mathscr{B}_{\text{Doping}}= & \frac{\Phi_{0}}{2\pi} \frac{\mu \partial_{x}g}{2ta \sin(K_{F}a)},
\end{align}
\enni for the inter-LL spacing and pseudo-magnetic field
respectively. Note that $\Phi_{0}=hc/e=4.12\times 10^{5}$~T\AA$^2$
is the quantum of flux.

\end{document}